\documentclass[a4paper,conference]{IEEEtran}
\usepackage{color,graphicx}
\usepackage[applemac]{inputenc}
\usepackage{epsfig}
\usepackage{latexsym}
\usepackage{amsfonts}
\usepackage{amssymb}
\usepackage{amsmath}
\usepackage{amsbsy}
\usepackage{color,graphicx}
\usepackage{multirow}
\usepackage{float}
\usepackage[english]{babel}
\usepackage{url}
\usepackage{subfigure}

\usepackage{cite}

\newcommand{\ul}{\underline}
\newcommand{\uul}[1]{\underline{\underline{#1}}}

\begin{document}
%
%
\title{Joint Estimation of the Time Delay and the Clock Drift and Offset Using UWB signals}
%
%
%
\author{\authorblockN{A. Mallat and L. Vandendorpe}
\authorblockA{Universit\'e catholique de Louvain\\
ICTEAM Institute\\
Place du Levant 2, B-1348 Louvain-la-Neuve, Belgium\\
Email: \{Achraf.Mallat, Luc.Vandendorpe\}@uclouvain.be}
}
\maketitle
\begin{abstract}
We consider two transceivers, the first with perfect clock and the second with imperfect clock. We investigate the joint estimation of the delay between the transceivers and the offset and the drift of the imperfect clock.
We propose a protocol for the synchronization of the clocks.
We derive some empirical estimators for the delay, the offset and the drift, and compute the Cramer-Rao lower bounds and the joint maximum likelihood estimator of the delay and the drift.
We study the impact of the protocol parameters and the time-of-arrival estimation variance on the achieved performances. 
We validate some theoretical results by simulation.
\end{abstract}


%
\IEEEpeerreviewmaketitle


\section{Introduction}\label{intro_sec}

Highly accurate positioning can be performed by employing the time-of-arrival (TOA) technique if impulse-radio (IR) ultra wideband (UWB) signals \cite{fcc,ec1,ec2} are transmitted.

\smallskip

However, one of the main challenges facing the realization of UWB-based positioning systems is the need of synchronization among all the network transceivers if the TOA technique is used and among the reference nodes if the time-difference-of-arrival (TDOA) is used. 
Synchronization can be accomplished by using high-precision clocks which seems to be impractical due to the required high-cost. 
To overcome this problem two-way ranging strategies can be used as proposed in \cite{Lee} and adopted in the IEEE802.15.4a standard \cite{IEEE802.15.4a,GeziciIEEE,dardari3}. 
Two-way ranging can mitigate the effects of the offset between clocks. However, the impact of clock drift is still present and causes non-negligible errors when the waited time at the receiver side is relatively long \cite{Kim,Xing,Jiang,dardari3}.

\smallskip

The effects of clock drift on TOA estimation accuracy is evaluated in many works where a wide variety of two-way protocols are proposed to reduce as much as possible the impact of the drift \cite{Denis,Jiang,Kim,Shimizu,Sivrikaya,Xing,Zhen}. 
However the problem of joint delay and clock offset and drift estimation is not investigated, or is investigated but without taking into account the primary impact of TOA estimation errors. Even when TOA estimation errors are considered they are either considered in simulation only, or are considered in the proposed model but the proposed estimators are not optimal.

\smallskip

In this paper we consider two transceivers, one equipped with a perfect clock and one equipped with an imperfect clock.  
We investigate the joint estimation of the time delay between the two transceivers and the offset and the drift of the imperfect clock. 
We propose a system model taking into account the TOA estimation errors at both transceivers.
We compute the Cramer-Rao lower bounds (CRLB) for the joint estimation of the delay and the drift and derive the joint maximum likelihood estimator (MLE).
Also, we propose some empirical estimators for the delay, the clock offset and the drift. 
The impact of the different parameters of the protocol and the TOA estimation variance on the proposed estimators is examined. 
The theoretical results are validated by simulation.
The approach followed in this paper can be extended to derive the CRLBs and the joint MLE for many synchronization protocols under different assumptions.

\smallskip

The rest of the paper is organized as follows. 
In Sec. \ref{model_sec}, we describe the system model. 
In Sec. \ref{protocol_sec}, we present the estimation protocol. 
In Sec. \ref{empirical_sec}, we propose an empirical algorithm.
In Sec. \ref{mle_sec}, we derive the CRLBs and the MLE.
In Sec. \ref{results_sec}, we show and discuss some numerical results.


\section{System model}\label{model_sec}

As mentioned above, we describe in this section our system model.
Let us consider two transceivers $\text{Tr}$ and $\text{Tr}'$ equipped with two clocks $\text{Ck}$ and $\text{Ck}'$, respectively, and assume that:
\begin{enumerate}
	\item The clock $\text{Ck}$ is perfect whereas the clock $\text{Ck}'$ suffers from a drift and an offset.
	\item The time delay $\tau$ between $\text{Tr}$ and $\text{Tr}'$ (i.e. $\tau$ is the time spent by a signal transmitted by $\text{Tr}$ to reach $\text{Tr}'$) is constant. Therefore, if $\text{Tr}$ and $\text{Tr}'$ communicate through free space (resp. a cable) then the distance (resp. the cable length) should be constant.
	In multipath channels, $\tau$ is proportional to the length of the detected path (not necessarily the direct one).
\end{enumerate}
The local time $t'$ of $\text{Ck}'$ can be written with respect to (w.r.t.) the true time $t$ (local time of $\text{Ck}$) as:
\begin{equation}\label{tp_eq}
	t' = \alpha t+\gamma = (1+\nu)t+\gamma
\end{equation}
where $\nu=\alpha-1$ (a coefficient) and $\gamma$ (in seconds) denote the drift and the offset of $\text{Ck}'$, respectively. The drift is often expressed in terms of parts-per-million (ppm); it is defined as the maximum number of extra or missed clock counts over a total of $10^6$ counts. The drift as defined in \eqref{tp_eq} is obtained from that in ppm $\nu_{\text{ppm}}$ by $\nu=\nu_{\text{ppm}}10^{-6}$. We assume in this paper that $\nu$ can be positive or negative. 

\smallskip

\begin{figure}
  \centering
  \includegraphics[width=7cm]{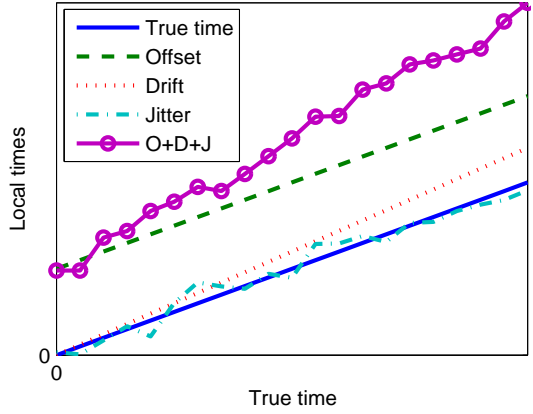}
  \caption{True time (solid line), local time with offset (dashed line), local time with drift (dotted line), local time with jitter (dash-dotted line), and local time with offset, drift and jitter (line with circles).}
  \label{off_drift_jitter_fig}
\end{figure}

Similarly to \cite{dardari3,Denis,Jiang,Kim,Sivrikaya,Xing} the problem of clock jitter is not included in our model for simplicity reasons. The jitter denotes the instantaneous fluctuations around the average local time described in \eqref{tp_eq}. In Fig. \ref{off_drift_jitter_fig}, we illustrate the lines representing the true time (solid line), a local time with an offset (dashed line), a local time with drift (dotted line), a local with jitter (dash-dotted line), and a local time with all the mentioned imperfections (line with circles).

\smallskip

In the next sessions we propose a protocol and some algorithms to synchronize $\text{Tr}$ and $\text{Tr}'$ and to estimate the time delay between them.

\section{Estimation protocol}\label{protocol_sec}

In this section we describe our estimation protocol. ``Protocol" stands for the consecutive steps to be followed by $\text{Tr}$ and $\text{Tr}'$ in order to obtain the observation carrying the information about the unknown parameters.
After proposing our protocol we realized that a similar protocol has already been proposed in \cite{Kim}.
The main contribution in this work is the derivation of the CRLBs and the estimation algorithms presented in the next sections rather than the protocol described here.

\smallskip

Let us present the protocol:
\begin{enumerate}
	\item $\text{Tr}'$ sends a signal to $\text{Tr}$ at the ``time of departure" (TOD) $t_D'$ (TOD w.r.t. $\text{Ck}'$); using \eqref{tp_eq}, we can write $t_D'$ w.r.t. the true TOD $t_D$ as:
\begin{eqnarray}
	t_D' = \alpha t_D+\gamma. \nonumber
\end{eqnarray}
	The transmitted signal arrives to $\text{Tr}$ at the true ``time of arrival" (TOA)
\begin{eqnarray}
	t_A = t_D+\tau = \frac{t_D'-\gamma}{\alpha}+\tau. \nonumber
\end{eqnarray}
	\item $\text{Tr}$ estimates $t_A$; denote by $\hat{t}_A$ the estimated TOA w.r.t. the perfect clock. We can write $\hat{t}_A$ as:
\begin{eqnarray}
	\hat{t}_A &=& t_A+\epsilon_A = \frac{t_D'-\gamma}{\alpha}+\tau+\epsilon_A \label{tAest_eq}
\end{eqnarray}
	where $\epsilon_A$ denotes the estimation error.
	\item $\text{Tr}$ waits for the durations $\delta_1,\cdots,\delta_N$ (known in advance by $\text{Tr}'$) before sending $N$ reply signals to $\text{Tr}'$. 
	We will see later in Sec. \ref{errorfree_subsec} that $N$ cannot be lower than two.
	The $n$th signal is transmitted at the true ``time of departure after waiting" (TOW) 
\begin{eqnarray}
	t_{W,n} = \hat{t}_A+\delta_n = \frac{t_D'-\gamma}{\alpha}+\tau+\delta_n+\epsilon_A \label{tWn_eq}
\end{eqnarray}
	and arrives to $\text{Tr}'$ at the true ``time of return" (TOR)
\begin{eqnarray}
	t_{R,n} = t_{W,n}+\tau = \frac{t_D'-\gamma}{\alpha}+2\tau+\delta_n+\epsilon_A \nonumber
\end{eqnarray}
which corresponds w.r.t. $\text{Ck}'$ to
\begin{align}
	t_{R,n}' &= \alpha t_{R,n}+\gamma  = \alpha(t_{W,n}+\tau)+\gamma \nonumber\\
				&= t_D'+\alpha(2\tau+\delta_n)+\alpha\epsilon_A. \nonumber
\end{align}
%
	\item $\text{Tr}'$ estimates $t_{R,n}'$; denote by $\hat{t}_{R,n}'$ the estimated TOR w.r.t. $\text{Ck}'$; $\hat{t}_{R,n}'$ can be written as:
\begin{align}
	\hat{t}_{R,n}' &= t_{R,n}'+\epsilon_{R,n}' = \alpha(t_{W,n}+\tau)+\gamma+\epsilon_{R,n}' \label{tRnpest_eq}\\
				&= t_D'+\alpha(2\tau+\delta_n)+\alpha\epsilon_A+\epsilon_{R,n}' \label{tRnpest1_eq}
\end{align}
	where $\epsilon_{R,n}'$ denotes the estimation error w.r.t. $\text{Ck}'$.
	\item $\text{Tr}'$ proceeds to the estimation of the unknown parameters $\alpha$, $\gamma$ and $\tau$ by making use of the protocol parameters $\delta_1,\cdots,\delta_N$, the estimated TOA $\hat{t}_A$ and TORs $\hat{t}_{R,1}',\cdots,\hat{t}_{R,N}'$, and the distributions of the estimation errors $\epsilon_A,\epsilon_{R,1}',\cdots,\epsilon_{R,N}'$ (possible to be estimated jointly with the TOA and the TORs).
\end{enumerate}

To be able to estimate the clock offset $\gamma$, the estimated TOA $\hat{t}_A$ should be contained in the reply signals sent by $\text{Tr}$ to $\text{Tr}'$. Otherwise, $\text{Tr}'$ can only estimate the time delay $\tau$ and the clock drift $\alpha$ (from the estimated TORs $\hat{t}_{R,1}',\cdots,\hat{t}_{R,N}'$).

\smallskip

It can be shown \cite{mcaulay3} that in the presence of an additive white Gaussian noise (AWGN), the MLE of the TOA is unbiased at sufficiently high signal-to-noise ratios (SNR), follows a normal distribution and achieves the CRLB. At low and medium SNRs, the MLE is no longer Gaussian. Let us assume that:
\begin{eqnarray}
	\epsilon_A &\sim& \mathcal{N}(0,\sigma_A^2) \label{pdfA_eq}\\
	\epsilon_{R,n}' &\sim& \mathcal{N}(0,\sigma_{R'}^2=\alpha^2\sigma_R^2\approx\sigma_R^2), \forall n \label{pdfR_eq}
\end{eqnarray}
where $\mathcal{N}(\mu_{\mathcal{N}},\sigma_{\mathcal{N}}^2)$ denotes the normal distribution of mean $\mu_{\mathcal{N}}$ and variance $\sigma_{\mathcal{N}}^2$, $\sigma_A^2$ the variance of $\hat{t}_A$ w.r.t. a perfect clock, and $\sigma_{R'}^2$ and $\sigma_R^2$ the variances of $\hat{t}_{R,n}'$ w.r.t. an imperfect and a perfect clock respectively; $\sigma_{R'}^2$ is approximated by $\sigma_R^2$ to make the covariance matrices in Secs. \ref{empirical_sec}, \ref{mle_sec} independent of the unknown parameters to estimate (valid assumption because $\alpha\approx1$ and because $\sigma_R$ is much smaller than $\tau$ and $\delta_n$).
In \eqref{pdfR_eq}, $\sigma_R^2$ is the same $\forall n$ because the reply signals sent by $\text{Tr}$ to $\text{Tr}'$ have all the same energy; $\sigma_A^2$ is not assumed equal to $\sigma_R^2$ because the signals transmitted by $\text{Tr}$ and $\text{Tr}'$ do not necessarily have the same energy.

\smallskip

At sufficiently high SNRs, both $\sigma_A^2$ and $\sigma_R^2$ can be computed from the expression of the CRLB for time estimation given by \cite{mallat1,mallat2009mpc}
\begin{eqnarray}
	c_T = \frac{1}{\rho\beta^2} \nonumber
\end{eqnarray}
where $\rho$ and $\beta^2$ denote the SNR and the mean quadratic bandwidth of the transmitted signal ($\beta$ is also called effective bandwidth) respectively.
For a signal occupying the whole UWB band authorized by the US federal commission of communications (FCC) \cite{fcc} (central frequency of $6.85$ GHz and bandwidth of $7.5$ GHz so $\beta=45.14$ GHz) we have $\sqrt{c_T}=7$ ps (resp. $0.7$ ps) at $\rho=10$ dB (resp. $30$ ps).


\section{Empirical algorithm}\label{empirical_sec}

In this section we propose an empirical algorithm for the estimation of the time delay and the clock offset and drift. 
We consider in Sec. \ref{errorfree_subsec} the case where the estimation errors $\epsilon_A$ in \eqref{tAest_eq} and $\epsilon_{R,n}'$ in \eqref{tRnpest1_eq} are null (i.e $\hat{t}_A$ and $\hat{t}_{R,n}'$ correctly estimated) and present in Sec. \ref{empirical_subsec} the proposed algorithm.

\smallskip

Note that an optimal estimator should treat the entire available observation. Accordingly, if $\hat{t}_A$ is known (resp. unknown) by $\text{Tr}'$ then $\alpha$, $\gamma$ and $\tau$ (resp. $\alpha$ and $\tau$) should be jointly estimated by maximizing the likelihood function relative to $\hat{t}_A$ and $\hat{t}_{R,1}',\cdots,\hat{t}_{R,N}'$ (resp. $\hat{t}_{R,1}',\cdots,\hat{t}_{R,N}'$).


\subsection{Error-free case}\label{errorfree_subsec}

As mentioned above we assume here that $\epsilon_A$ in \eqref{tAest_eq} and $\epsilon_{R,n}'$ in \eqref{tRnpest1_eq} are null.

\smallskip

To find $\alpha$ and $\tau$ from \eqref{tRnpest1_eq}, we need at least two equations. So by taking $N=2$ we can write
\begin{eqnarray}
	\left\{\begin{array}{rcl} 
				\hat{t}_{R,1}' &=& t_D'+\alpha(2\tau+\delta_1) \\
				\hat{t}_{R,2}' &=& t_D'+\alpha(2\tau+\delta_2) 
	\end{array}\right. \nonumber
\end{eqnarray}
so $\alpha$ and $\tau$ can be expressed as ($n=1,2$):
\begin{eqnarray}
	\alpha &=& \frac{\hat{t}_{R,2}'-\hat{t}_{R,1}'}{\delta_2-\delta_1} \label{alpha_eq}\\
	\tau &=& \frac{\hat{t}_{R,n}'-t_D'-\alpha\delta_n}{2\alpha}. \label{tau_eq}
\end{eqnarray}
Note that $\tau$ is also given by 
\begin{eqnarray}
	\tau = \frac{\delta_2(\hat{t}_{R,1}'-t_D')-\delta_1(\hat{t}_{R,2}'-t_D')}{2(\hat{t}_{R,2}'-\hat{t}_{R,1}')}. \nonumber
\end{eqnarray}
However, we prefer the expression in \eqref{tau_eq} because it will be used later in Sec. \ref{empirical_subsec} in the proposed algorithm.

\smallskip

If we assume that $\hat{t}_A$ is know by $\text{Tr}'$, then $\gamma$ can be expressed from \eqref{tAest_eq}, \eqref{tWn_eq} and \eqref{tRnpest_eq} as ($n=1,2$):
\begin{align}
	\gamma &= t_D'-\alpha(\hat{t}_A-\tau) \label{gamma_eq}\\
				&= \hat{t}_{R,n}'-\alpha(\hat{t}_A+\delta_n+\tau). \label{gamma1_eq}
\end{align}

Hence, $N=2$ is sufficient to obtain the exact values of the unknown parameters in the error-free case. In the presence of errors, $N=2$ is also sufficient to perform the estimation; however, estimation performance can be improved by increasing the number of observations.


\subsection{The proposed algorithm}\label{empirical_subsec}

Many empirical estimators for $\alpha$, $\tau$ and $\gamma$ can be proposed based on the equations established in Sec. \ref{protocol_sec}. However, it will suffice to investigate one estimator only as an example. The main goal is to compare the performances of an empirical estimator with the performances of the optimal estimator considered in Sec. \ref{mle_sec}.

\smallskip

From \eqref{alpha_eq}, we can generate the following $N-1$ estimates of $\alpha$ ($n=1,\cdots,N-1$):
\begin{eqnarray}
	\hat{\alpha}_{n,1} = \frac{\hat{t}_{R,n+1}'-\hat{t}_{R,1}'}{\delta_{n+1}-\delta_1}. \label{alphaestn_eq}
\end{eqnarray}
Let $\ul{\hat{\alpha}_1}=(\hat{\alpha}_{1,1},\cdots,\hat{\alpha}_{N-1,1})^T$ with $^T$ denoting the transpose operator. 
By considering $\ul{\hat{\alpha}_1}$ as the observation carrying the information on $\alpha$, the log-likelihood function for the estimation of $\alpha$ can be written from \eqref{pdfR_eq} and \eqref{alphaestn_eq} as:
\begin{eqnarray}
	\Lambda_{\ul{\hat{\alpha}_1}} = -\frac{1}{2}\left(\ul{\hat{\alpha}_1}-\ul{\mu}_{\ul{\hat{\alpha}_1}}\right)^T\uul{\Omega}_{\ul{\hat{\alpha}_1}}^{-1}\left(\ul{\hat{\alpha}_1}-\ul{\mu}_{\ul{\hat{\alpha}_1}}\right) \nonumber
\end{eqnarray}
where
\begin{eqnarray}
	\ul{\mu}_{\ul{\hat{\alpha}_1}} &=& \alpha\ul{1}_{N-1} \nonumber\\
	\uul{\Omega}_{\ul{\hat{\alpha}_1}} &=& \left(\omega_{m,n}\right)_{m,n=1,\cdots,N-1} \nonumber
\end{eqnarray}
denote the mean and the covariance matrix of $\ul{\hat{\alpha}_1}$ with $\ul{1}_{N-1}$ being a vector of $N-1$ elements equal to one, and
\begin{eqnarray}
	\omega_{m,n} = \left\{\begin{array}{ll} \frac{2\sigma_R^2}{(\delta_n-\delta_1)^2} & m=n \\ 
															\frac{\sigma_R^2}{(\delta_m-\delta_1)(\delta_n-\delta_1)} & m\neq n. \end{array}\right. \nonumber
\end{eqnarray}
The MLE $\hat{\alpha}_1$ (w.r.t. to the observation $\ul{\hat{\alpha}_1}$) of $\alpha$ consists on maximizing the log-likelihood function $\Lambda_{\ul{\hat{\alpha}_1}}$. 
The partial derivative of $\Lambda_{\ul{\hat{\alpha}_1}}$ w.r.t. $\alpha$ can be written as:
\begin{align}
	\frac{\partial\Lambda_{\ul{\hat{\alpha}_1}}}{\partial\alpha} &=
							\left(\frac{\partial\ul{\mu}_{\ul{\hat{\alpha}_1}}}{\partial\alpha}\right)^T
							\uul{\Omega}_{\ul{\hat{\alpha}_1}}^{-1}\left(\ul{\hat{\alpha}}-\ul{\mu}_{\ul{\hat{\alpha}_1}}\right) \nonumber\\
				&= \ul{1}_{N-1}^T\uul{\Omega}_{\ul{\hat{\alpha}_1}}^{-1}\ul{\hat{\alpha}_1}
							-\alpha\ul{1}_{N-1}^T\uul{\Omega}_{\ul{\hat{\alpha}_1}}^{-1}\ul{1}_{N-1}. \nonumber
\end{align}
By equating $\frac{\partial\Lambda_{\ul{\hat{\alpha}_1}}}{\partial\alpha}$ to zero we can express $\hat{\alpha}_1$ as:
\begin{align}
	\hat{\alpha}_1 = \frac{\ul{a}^T\ul{\hat{\alpha}_1}}{A} \label{alphaml1_eq}
\end{align}
with
\begin{eqnarray}
	\ul{a}^T &=& \ul{1}_{N-1}^T\uul{\Omega}_{\ul{\hat{\alpha}_1}}^{-1} \nonumber\\
	A &=& \ul{1}_{N-1}^T\uul{\Omega}_{\ul{\hat{\alpha}_1}}^{-1}\ul{1}_{N-1}. \nonumber
\end{eqnarray}
We can see from \eqref{alphaml1_eq} that $\hat{\alpha}_1$ follows a normal distribution with a mean and a variance respectively given by
\begin{eqnarray}
	\mu_{\hat{\alpha}_1} &=& \frac{\ul{a}^T\mu_{\ul{\hat{\alpha}_1}}}{A} = \alpha \nonumber\\
	\sigma^2_{\hat{\alpha}_1} &=& \frac{\ul{a}^T\uul{\Omega}_{\ul{\hat{\alpha}_1}}\ul{a}}{A^2} = \frac{1}{A}. \label{sigma2alpha1_eq}
\end{eqnarray}
Our estimator is thus unbiased.
We have considered $\hat{\alpha}_1$ as empirical because $\ul{\hat{\alpha}_1}$ is not necessarily a sufficient statistic. 

\smallskip

From \eqref{tau_eq} and \eqref{alphaml1_eq}, we can generate the following $N$ estimates of $\tau$ ($n=1,\cdots,N$):
\begin{eqnarray}
	\hat{\tau}_{n,1} = \frac{\hat{t}_{R,n}'-t_D'-\hat{\alpha}_1\delta_{n}}{2\hat{\alpha}_1}. \label{tauestn_eq}
\end{eqnarray}
The variance of $\hat{\tau}_{n,1}$ is not the same $\forall n$ due to the term $\hat{\alpha}_1\delta_{n}$ (the variance of $\hat{\alpha}_1$ is proportional to $\sigma_R^2$); we recall that the variance of $\hat{t}_{R,n}'$ is equal to $\sigma_A^2+\sigma_R^2$, $\forall n$.
From $\hat{\tau}_{n,1}$ in \eqref{tauestn_eq}, we propose the following estimator of $\tau$:
\begin{eqnarray}
	\hat{\tau}_1 = \frac{\ul{1}_{N}^T\ul{\hat{\tau}_1}}{N} \label{tauml1_eq}
\end{eqnarray}
where $\ul{\hat{\tau}_1}=(\hat{\tau}_{1,1},\cdots,\hat{\tau}_{N,1})^T$.

\smallskip

From \eqref{gamma1_eq}, \eqref{alphaml1_eq} and \eqref{tauml1_eq}, we can generate the following $N$ estimates of $\gamma$ ($n=1,\cdots,N$):
\begin{eqnarray}
	\hat{\gamma}_{n,1} = \hat{t}_{R,n}'-\hat{\alpha}_1\left(\hat{t}_A+\delta_{n}+\hat{\tau}_1\right). \label{gammaestn_eq}
\end{eqnarray}
From $\hat{\gamma}_{n,1}$ in \eqref{gammaestn_eq}, we propose the following estimator of $\gamma$:
\begin{eqnarray}
	\hat{\gamma}_{11} = \frac{\ul{1}_{N}^T\ul{\hat{\gamma}_1}}{N} \label{gammaml11_eq}
\end{eqnarray}
where $\ul{\hat{\gamma}_1}=(\hat{\gamma}_{1,1},\cdots,\hat{\gamma}_{N,1})^T$.
Another estimator can be directly proposed from \eqref{gamma_eq} as:
\begin{eqnarray}
	\hat{\gamma}_{12} = t_D'-\hat{\alpha}_1(\hat{t}_A-\hat{\tau}_1). \label{gammaml12_eq}
\end{eqnarray}

Note that the exact means and variances of $\hat{\tau}_{n,1}$ in \eqref{tauestn_eq} and $\hat{\gamma}_{n,1}$ in \eqref{gammaestn_eq} are not easy to express because $\hat{\tau}_{n,1}$ is the ratio of two random variables. However, the asymptotic statistics are possible to compute. Nevertheless, we did not calculate them here for the sake of conciseness.


\section{CRLBs and joint MLE}\label{mle_sec}

In this section we derive the CRLBs for the joint estimation of the time delay $\tau$ and the clock drift $\alpha$ based on the estimated TORs $\hat{t}_{R,n}'$ in \eqref{tRnpest1_eq}. We compute the joint MLE of $\alpha$ and $\tau$ and propose two empirical estimators for the clock offset $\gamma$.

\smallskip

Let:
\begin{align}
	\ul{X} = \ul{\hat{t}_{R}'}-t_D'\ul{1}_N \nonumber
\end{align}
where $\ul{\hat{t}_{R}'}=(\hat{t}_{R,1}',\cdots,\hat{t}_{R,N}')^T$.
The log-likelihood function for the joint estimation of $\alpha$ and $\tau$ can be written from \eqref{tRnpest1_eq}--\eqref{pdfR_eq} as:
\begin{align}
	\Lambda_{\ul{X}} =
				-\frac{1}{2}\left(\ul{X}-\ul{\mu}_{\ul{X}}\right)^T\uul{\Omega}_{\ul{X}}^{-1}\left(\ul{X}-\ul{\mu}_{\ul{X}}\right) \nonumber
\end{align}
where
\begin{eqnarray}
	\mu_{\ul{X}} &=& \alpha(2\tau\ul{1}_N+\ul{\delta}) \label{muX_eq}\\
	\uul{\Omega}_{\ul{X}} &=& \left(\omega_{m,n}\right)_{m,n=1,\cdots,N} \label{CovX_eq}
\end{eqnarray}
respectively denote the mean and the covariance matrix of $\ul{X}$ with $\ul{\delta}=(\delta_1\cdots\delta_N)^T$ and
\begin{eqnarray}
	\omega_{m,n} = \left\{\begin{array}{ll} \sigma_A^2+\sigma_R^2 & m=n \\ 
															\sigma_A^2 & m\neq n. \end{array}\right. \nonumber
\end{eqnarray}


\subsection{CRLBs for the joint estimation of $\alpha$ and $\tau$}\label{crlb_subsec}

The CRLB for the estimation of a parameter gives the lowest variance achievable by an unbiased estimator. Denote by $\mathbb{E}$ the expectation operator. The CRLBs of $\alpha$ and $\tau$ are \cite{kay} the diagonal elements of the inverse of the Fisher information matrix (FIM) given by
\begin{align}
	 	\mathcal{\uul{F}}_{\ul{X}} = \left(\begin{array}{cc} 
	 							f_{\alpha,\alpha} & f_{\alpha,\tau} \\
	 							f_{\tau,\alpha} & f_{\tau,\tau} \end{array} \right) \nonumber
\end{align}
where
\begin{align}
	 	f_{\theta,\theta'} = -\mathbb{E}\left\{\frac{\partial^2\Lambda_{\ul{X}}}{\partial\theta\partial\theta'}\right\} 
	 				= \frac{\partial\ul{\mu}_{\ul{X}}^T}{\partial\theta'} \uul{\Omega}_{\ul{X}}^{-1} \frac{\partial\ul{\mu}_{\ul{X}}}{\partial\theta} = f_{\theta',\theta} \nonumber
\end{align}
with $\theta,\theta'\in\{\alpha,\tau\}$ and
\begin{eqnarray}
	\frac{\partial\ul{\mu}_{\ul{X}}}{\partial\alpha} &=& 2\tau\ul{1}_N+\ul{\delta} \label{dmuXalpha_eq}\\
	\frac{\partial\ul{\mu}_{\ul{X}}}{\partial\tau} &=& 2\alpha\ul{1}_N. \label{dmuXtau_eq}	
\end{eqnarray}
Hence,
\begin{eqnarray}
	 	f_{\alpha,\alpha} &=& 4\tau^2B+4\tau D+F \nonumber\\
	 	f_{\tau,\tau} &=& 4\alpha^2B \nonumber\\
	 	f_{\alpha,\tau} &=& 2\alpha(2\tau B+D) = f_{\tau,\alpha} \nonumber
\end{eqnarray}
where
\begin{eqnarray}
	B &=& \ul{1}^T_N\uul{\Omega}_{\ul{X}}^{-1}\ul{1}_N \nonumber\\
	D &=& \ul{1}^T_N\uul{\Omega}_{\ul{X}}^{-1}\ul{\delta} \nonumber\\
	F &=& \ul{\delta}^T\uul{\Omega}_{\ul{X}}^{-1}\ul{\delta}. \nonumber
\end{eqnarray}
The CRLBs of $\alpha$ and $\tau$ can respectively be expressed as:
\begin{eqnarray}
	 	c_{\alpha} &=& \frac{f_{\tau,\tau}}{f_{\alpha,\alpha}f_{\tau,\tau}-f_{\alpha,\tau}^2} 
	 				= \frac{B}{BF-D^2} \label{cpha_eq}\\
	 	c_{\tau} &=& \frac{f_{\alpha,\alpha}}{f_{\alpha,\alpha}f_{\tau,\tau}-f_{\alpha,\tau}^2} 
	 				= \frac{4\tau^2B+4\tau D+F}{4\alpha^2(BF-D^2)}. \label{ctau_eq}
\end{eqnarray}
We can show that $c_{\alpha}$ is a function of $\sigma_R^2$, $N$ and the variance of $\delta_n$ only. We can show as well that the term $4\tau^2B+4\tau D$ can be neglected in the expression of $c_{\tau}$ and that $\alpha$ can be approximated by $1$ so $c_{\tau}$ becomes a function of $\sigma_A^2$, $\sigma_R^2$, $N$ and the mean and the variance of $\delta_n$.


\subsection{Joint MLE of $\alpha$ and $\tau$}\label{mleb_subsec}

The MLE $(\hat{\alpha}_2,\hat{\tau}_2)$ of $(\alpha,\tau)$ consists on maximizing the log-likelihood function $\Lambda_{\ul{X}}$. Therefore, $(\hat{\alpha}_2,\hat{\tau}_2)$ can be obtained by equating the partial derivatives of $\Lambda_{\ul{X}}$ to zero:
\begin{eqnarray}
	\left\{\begin{array}{l} 
		\left.\frac{\partial\Lambda_{\ul{X}}}{\partial\alpha}
					\right|_{(\alpha,\tau)=(\hat{\alpha}_2,\hat{\tau}_2)}=0 \\
		\left.\frac{\partial\Lambda_{\ul{X}}}{\partial\tau}
					\right|_{(\alpha,\tau)=(\hat{\alpha}_2,\hat{\tau}_2)}=0
	\end{array}\right. \label{mlecondV_eq}
\end{eqnarray}
where
\begin{eqnarray}
	\frac{\partial\Lambda_{\ul{X}}}{\partial\theta} 
				= \frac{\partial\ul{\mu}_{\ul{X}}^T}{\partial\theta}\uul{\Omega}_{\ul{X}}^{-1}\left(\ul{X}-\ul{\mu}_{\ul{X}}\right) \nonumber \end{eqnarray}
with $\theta\in\{\alpha,\tau\}$.
Using \eqref{muX_eq}--\eqref{dmuXtau_eq}, we can write from \eqref{mlecondV_eq}:
\begin{eqnarray}
\left(2\hat{\tau}_2\ul{1}_N+\ul{\delta}\right)^T\uul{\Omega}_{\ul{X}}^{-1}\big[\ul{X}-\hat{\alpha}_2(2\hat{\tau}_2\ul{1}_N+\ul{\delta})\big]	&=& 0 \label{d0_1_eq}\\
	2\hat{\alpha}_2\ul{1}^T_N\uul{\Omega}_{\ul{X}}^{-1}\big[\ul{X}-\hat{\alpha}_2(2\hat{\tau}_2\ul{1}_N+\ul{\delta})\big] &=& 0. \label{d0_2_eq}
\end{eqnarray}
By taking account of \eqref{d0_2_eq}, \eqref{d0_1_eq} becomes:
\begin{eqnarray}
	\ul{\delta}^T\uul{\Omega}_{\ul{X}}^{-1}\big[\ul{X}-\hat{\alpha}_2(2\hat{\tau}_2\ul{1}_N+\ul{\delta})\big] = 0. \label{d0_3_eq}
\end{eqnarray}
After some manipulations, we can write \eqref{d0_2_eq} and \eqref{d0_3_eq} as:
\begin{eqnarray}
	C - 2\hat{\alpha}_2\hat{\tau}_2 B	- \hat{\alpha}_2 D &=& 0 \label{d0_4_eq}\\
	E - 2\hat{\alpha}_2\hat{\tau}_2 D	- \hat{\alpha}_2 F &=& 0 \label{d0_5_eq}
\end{eqnarray}
where
\begin{eqnarray}
	C &=& \ul{1}^T_N\uul{\Omega}_{\ul{X}}^{-1}\ul{X} \nonumber\\
	E &=& \ul{\delta}^T\uul{\Omega}_{\ul{X}}^{-1}\ul{X}. \nonumber
\end{eqnarray}
By solving the equation system in \eqref{d0_4_eq} and \eqref{d0_5_eq} we obtain the following expressions of $\hat{\alpha}_2$ and $\hat{\tau}_2$:
\begin{eqnarray}
	\hat{\alpha}_2 &=& \frac{BE-CD}{BF-D^2} = \ul{g}^T\ul{X} \label{alphaest_eq}\\
	\hat{\tau}_2 &=& \frac{CF-DE}{2(BE-CD)} = \frac{\ul{k}^T\ul{X}}{\ul{l}^T\ul{X}} \label{tauest_eq}
\end{eqnarray}
where
\begin{eqnarray}
	\ul{g}^T &=& \frac{\left(B\ul{\delta}^T-D\ul{1}^T_N\right)\uul{\Omega}_{\ul{X}}^{-1}}{BF-D^2} \nonumber\\
	\ul{k}^T &=& \left(F\ul{1}^T_N-D\ul{\delta}^T\right)\uul{\Omega}_{\ul{X}}^{-1} \nonumber\\
	\ul{l}^T &=& 2\left(B\ul{\delta}^T-D\ul{1}^T_N\right)\uul{\Omega}_{\ul{X}}^{-1}. \nonumber
\end{eqnarray}

In order to compute the statistics of our estimators we write $\ul{X}$, using \eqref{tRnpest1_eq}, in the expressions of $\hat{\alpha}_2$ and $\hat{\tau}_2$ as:
\begin{eqnarray}
	\ul{X} = \alpha(2\tau\ul{1}_N+\ul{\delta})+\ul{\epsilon_R} \label{X2_eq}
\end{eqnarray}
where $\ul{\epsilon_R}=\alpha\epsilon_A\ul{1}_N+(\epsilon_{R,1}'\cdots\epsilon_{R,N}')^T$; $\ul{\epsilon_R}$ is zero-mean and has the same covariance matrix as $\ul{X}$.
Then, 
\begin{eqnarray}
	\hat{\alpha}_2 &=& \ul{g}^T\big[\alpha(2\tau\ul{1}_N+\ul{\delta})+\ul{\epsilon_R}\big] 
				= \alpha + \ul{g}^T\ul{\epsilon_R} \label{alphaest1_eq}\\
	\hat{\tau}_2 &=& \frac{\ul{k}^T\big[\alpha(2\tau\ul{1}_N+\ul{\delta})+\ul{\epsilon_R}\big]}
							{\ul{l}^T\big[\alpha(2\tau\ul{1}_N+\ul{\delta})+\ul{\epsilon_R}\big]} 
				= \frac{\tau+\frac{\ul{k}^T\ul{\epsilon_R}}{2\alpha(BF-D^2)}}{1+\frac{\ul{l}^T\ul{\epsilon_R}}{2\alpha(BF-D^2)}} \label{tauest1_eq}\\
				 &\approx& \left[\tau+\frac{\ul{k}^T\ul{\epsilon_R}}{2\alpha(BF-D^2)}\right]
				 			\left[1-\frac{\ul{l}^T\ul{\epsilon_R}}{2\alpha(BF-D^2)}\right] \label{tauest2_eq}\\
				 &\approx& \tau + \frac{\left(\ul{k}^T-\tau\ul{l}^T\right)\ul{\epsilon_R}}{2\alpha(BF-D^2)}. \label{tauest3_eq}
\end{eqnarray}
We have obtained \eqref{tauest2_eq} from \eqref{tauest1_eq} by using the approximation $(1+\xi)^m\approx1+m\xi$ for $\xi<<1$, and \eqref{tauest3_eq} from \eqref{tauest2_eq} by neglecting the noise product (i.e. the noise of second order). 

\smallskip

We can see form \eqref{alphaest1_eq} that $\hat{\alpha}_2$ is unbiased and follows a normal distribution with a variance given by:
\begin{eqnarray}
	 \sigma^2_{\hat{\alpha}_2} = \ul{g}^T\uul{\Omega}_{\ul{X}}\;\ul{g} = \frac{B}{BF-D^2} = c_{\alpha}. \label{varalphaest_eq}
\end{eqnarray}
This result is very interesting because it shows that $\hat{\alpha}_2$ is efficient; it always achieves the CRLB.

\smallskip

Unlike $\hat{\alpha}_2$, $\hat{\tau}_2$ is biased and follows the distribution of the ratio of two correlated normal variables. The PDF of $\hat{\tau}_2$ can be computed by making use of the work in \cite{Marsaglia1964,Marsaglia2006} about the ratio of normal variables. For sufficiently high SNRs, $\hat{\tau}_2$ becomes, as can be observed from \eqref{tauest3_eq} unbiased and follows a normal distribution with a variance given by:
\begin{equation}
	 \sigma^2_{\hat{\tau}_2} = \frac{\left(\ul{k}^T-\tau\ul{l}^T\right)\uul{\Omega}_{\ul{X}}\Big(\ul{k}-\tau\ul{l}\Big)}{4\alpha^2(BF-D^2)^2}
	 			= \frac{4\tau^2B+4\tau D+F}{4\alpha^2(BF-D^2)} = c_{\tau}. \label{vartauest_eq}
\end{equation}
This result is very interesting as well because it shows that $\hat{\tau}_2$ is asymptotically efficient.


\subsection{Empirical estimators of $\gamma$}\label{algogamma_subsec}

Assume now that the TOA $\hat{t}_A$ is know by $\text{Tr}'$. The joint MLE of $\alpha$, $\gamma$ and $\tau$ consists in this case on maximizing the log-likelihood function corresponding to $\hat{t}_A$ and all $\hat{t}_{R,n}'$. This estimator is not investigated in this paper. In this subsection we propose two empirical estimators of $\gamma$ by making use of $\hat{\alpha}_2$ and $\hat{\tau}_2$ derived in the last subsection.

\smallskip

Similarly to the estimators in \eqref{gammaml11_eq} and \eqref{gammaml12_eq}, we propose the following two estimators:
\begin{eqnarray}
	\hat{\gamma}_{21} &=& \frac{\ul{1}_N^T\ul{\hat{\gamma}_2}}{N} \label{gamma21_eq}\\
	\hat{\gamma}_{22} &=& t_D'-\hat{\alpha}_2(\hat{t}_A-\hat{\tau}_2). \label{gamma22_eq}
\end{eqnarray}
where $\ul{\hat{\gamma}_2}=(\hat{\gamma}_{1,2},\cdots,\hat{\gamma}_{N,2})^T$ with
\begin{eqnarray}
	\hat{\gamma}_{n,2} = \hat{t}_{R,n}'-\hat{\alpha}_2\left(\hat{t}_A+\delta_n+\hat{\tau}_2\right). \nonumber
\end{eqnarray}


\section{Numerical results and discussion}\label{results_sec} 


In this section we discuss some numerical results. The main two goals are to evaluate our estimators and to study the impact of some parameters ($\sigma_A$, $\sigma_R$, $\delta_N$ and $N$) on the achieved performances.
Unfortunately, we cannot show all our results due to the lack of space.

\smallskip

Unless mentioned otherwise, we consider the following values in our simulations: 
$\nu_{\text{ppm}}=20$ ppm, $\gamma=1\,\mu\text{s}$, $\tau=100$ ns (which corresponds to a distance of 30 m), $\sigma_A=\sigma_R=0.1$ ns, $\delta_N=1$ ms, and $N=4$; $\delta_n$ is given by $\delta_n=\frac{n\delta_N}{N}$.
In our simulations the variances are obtained based on $10^4$ noise samples.

\smallskip

We denote by $\sigma_{\alpha,1}$, $\sigma_{\alpha,2}$, $\sigma_{\tau,1}$, $\sigma_{\tau,2}$, $\sigma_{\gamma,11}$, $\sigma_{\gamma,12}$, $\sigma_{\gamma,21}$ and $\sigma_{\gamma,22}$ the standard deviations (Stds) obtained by simulation of the estimators $\hat{\alpha}_1$ in \eqref{alphaml1_eq}, $\hat{\alpha}_2$ in \eqref{alphaest_eq}, $\hat{\tau}_1$ in \eqref{tauml1_eq}, $\hat{\tau}_2$ in \eqref{tauest_eq}, $\hat{\gamma}_{11}$ in \eqref{gammaml11_eq}, $\hat{\gamma}_{12}$ in \eqref{gammaml12_eq}, $\hat{\gamma}_{21}$ in \eqref{gamma21_eq} and $\hat{\gamma}_{22}$ in \eqref{gamma22_eq}, respectively,
by $\kappa_{\alpha,1}$ the Std of $\hat{\alpha}_1$ (square root of $\sigma^2_{\hat{\alpha}_1}$ in \eqref{sigma2alpha1_eq}),
and by $\kappa_{\alpha,2}$ and $\kappa_{\tau,2}$ the square roots of the CRLBs $c_{\alpha}$ in \eqref{cpha_eq} and $c_{\tau}$ in \eqref{ctau_eq}, respectively.

\smallskip

In Figs. \ref{alpha_sigA}--\ref{tau_N} we show the Stds for drift, offset and delay estimation, respectively, w.r.t. $\sigma_A$, $\sigma_R$, $\delta_N$ and $N$, respectively.


\subsection{Impact of $\sigma_A$}\label{sigA_subsec}

\begin{figure*}[ht]
	\begin{center}
		\subfigure[]{\includegraphics[width=6.1cm]{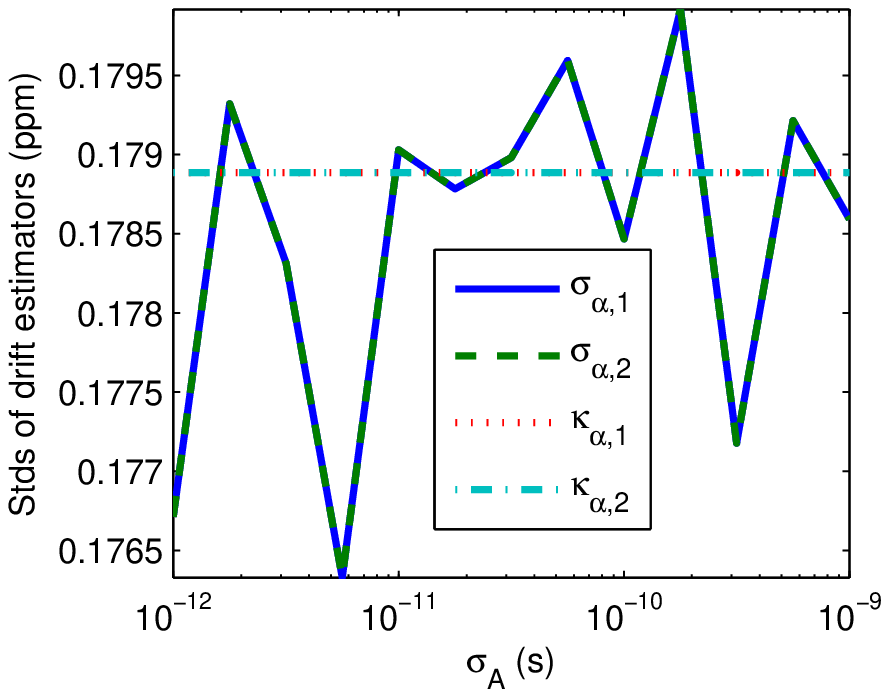} \label{alpha_sigA}} \hspace{-0.05\textwidth}
		\quad\subfigure[]{\includegraphics[width=6.1cm]{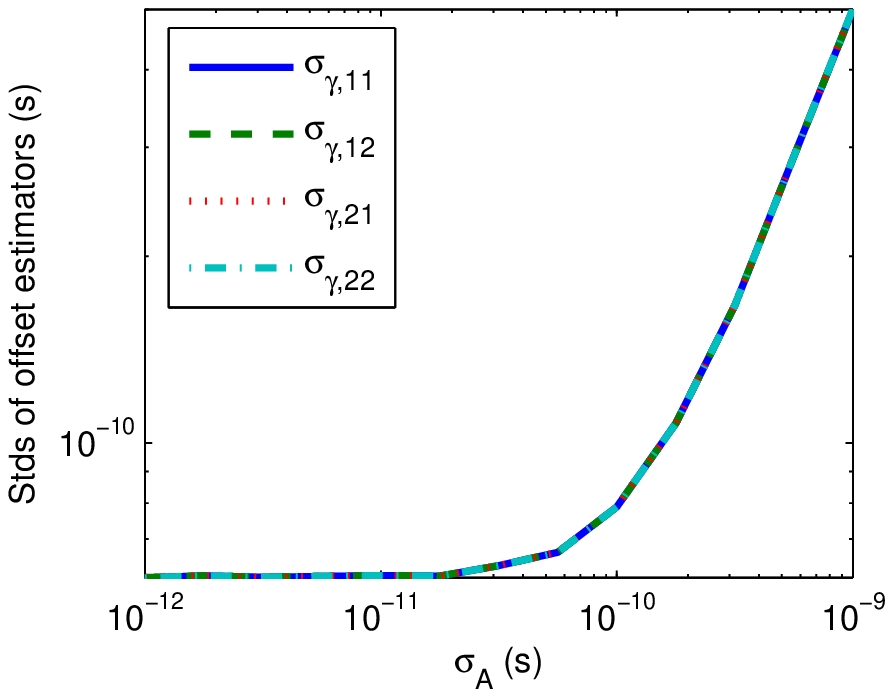} \label{gamma_sigA}} \hspace{-0.05\textwidth}
		\quad\subfigure[]{\includegraphics[width=6.1cm]{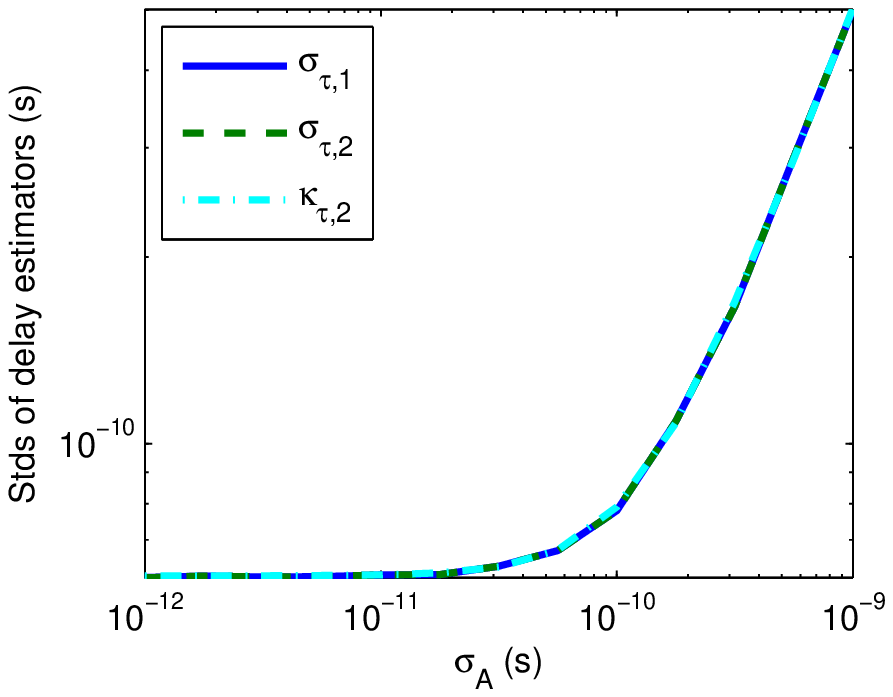} \label{tau_sigA}} 
		\quad\subfigure[]{\includegraphics[width=6.1cm]{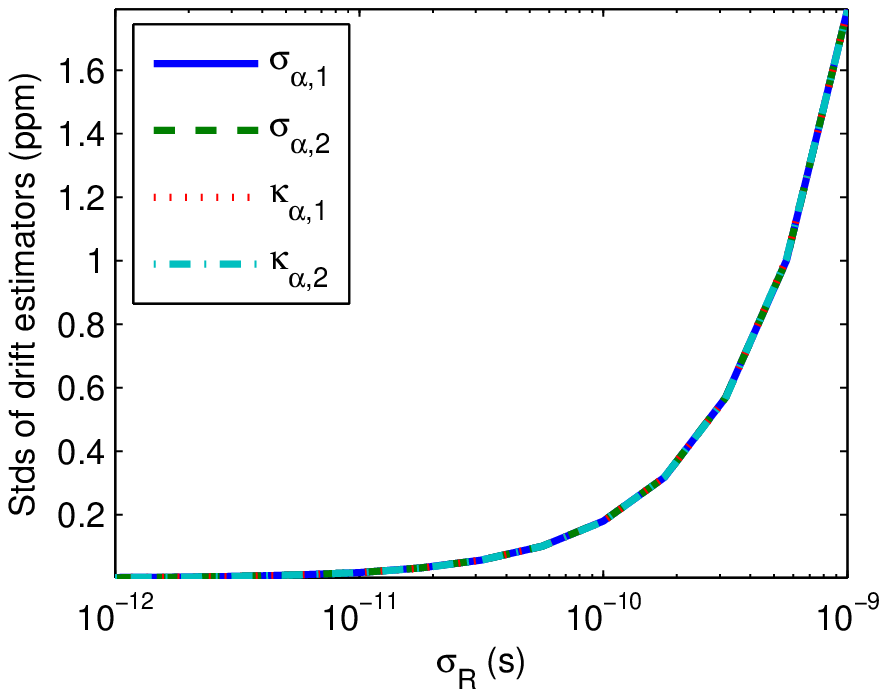} \label{alpha_sigR}} \hspace{-0.05\textwidth}
		\quad\subfigure[]{\includegraphics[width=6.1cm]{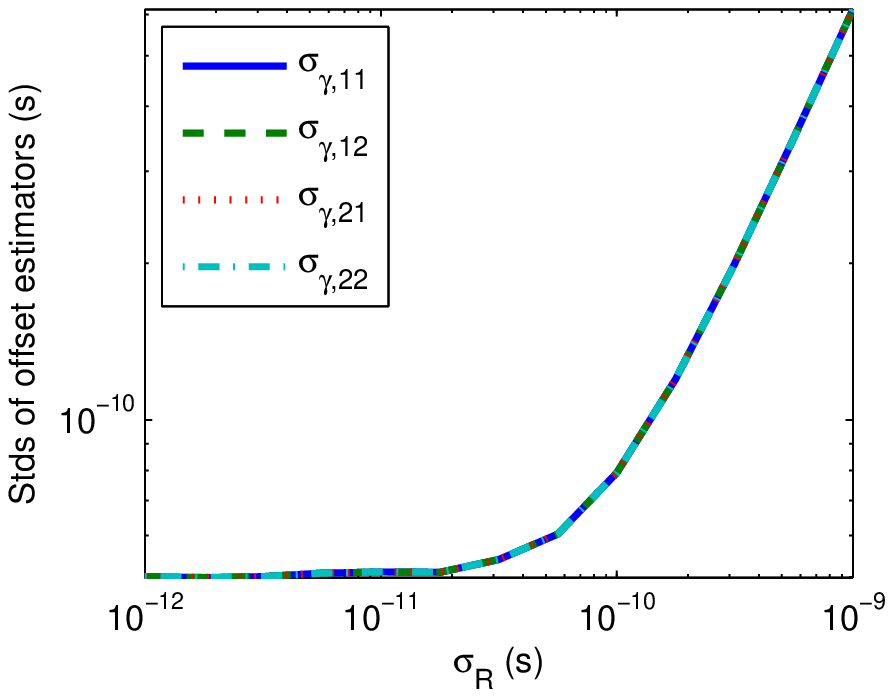} \label{gamma_sigR}} \hspace{-0.05\textwidth}
		\quad\subfigure[]{\includegraphics[width=6.1cm]{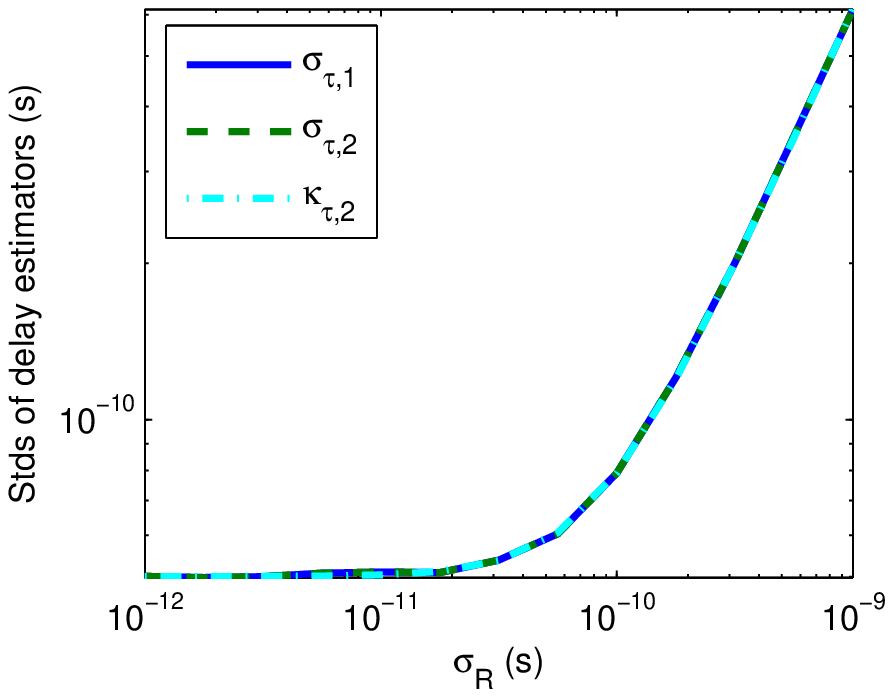} \label{tau_sigR}} 
		\quad\subfigure[]{\includegraphics[width=6.1cm]{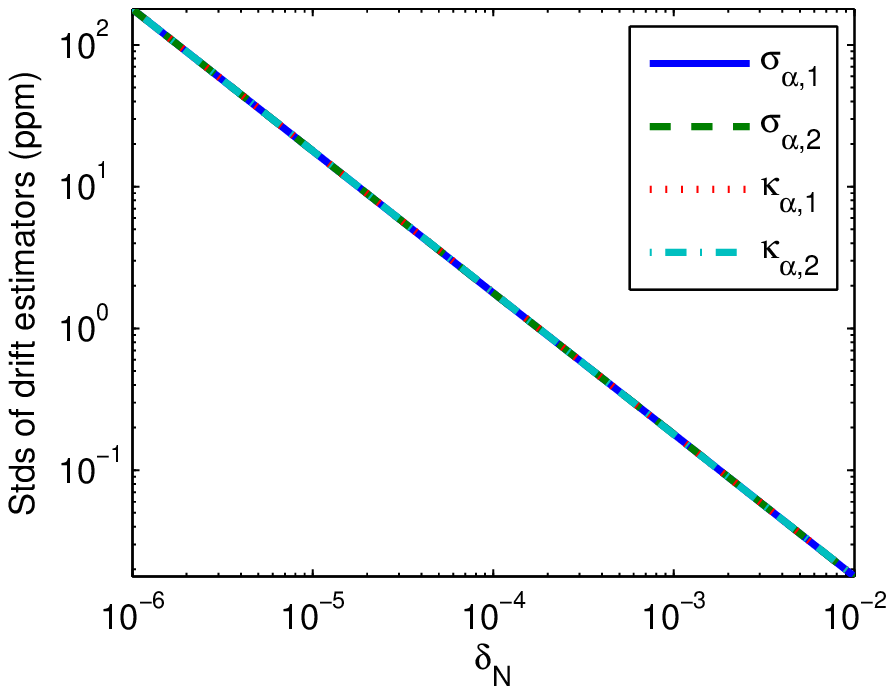} \label{alpha_delN}} \hspace{-0.05\textwidth}
		\quad\subfigure[]{\includegraphics[width=6.1cm]{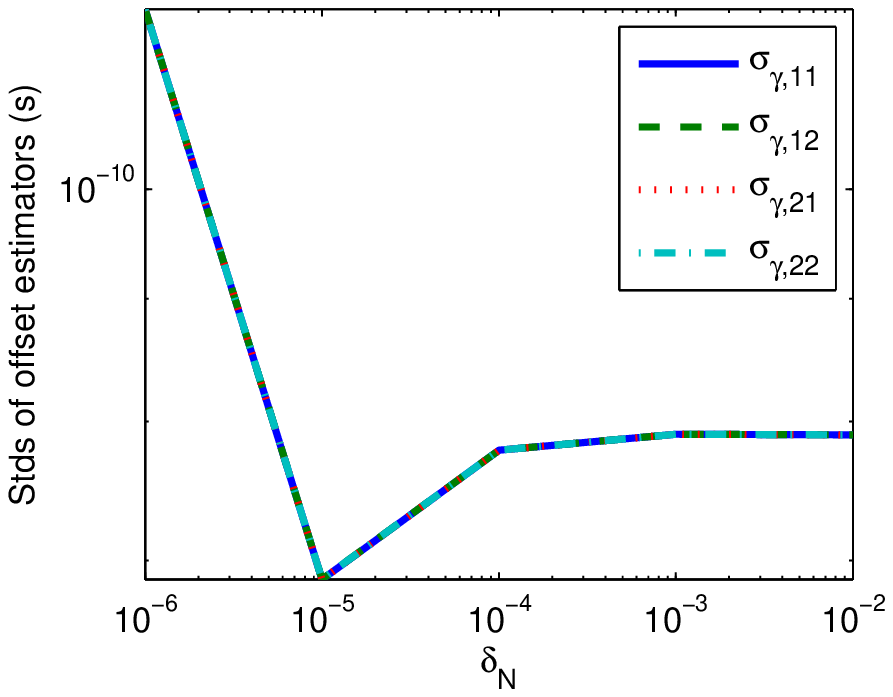} \label{gamma_delN}} \hspace{-0.05\textwidth}
		\quad\subfigure[]{\includegraphics[width=6.1cm]{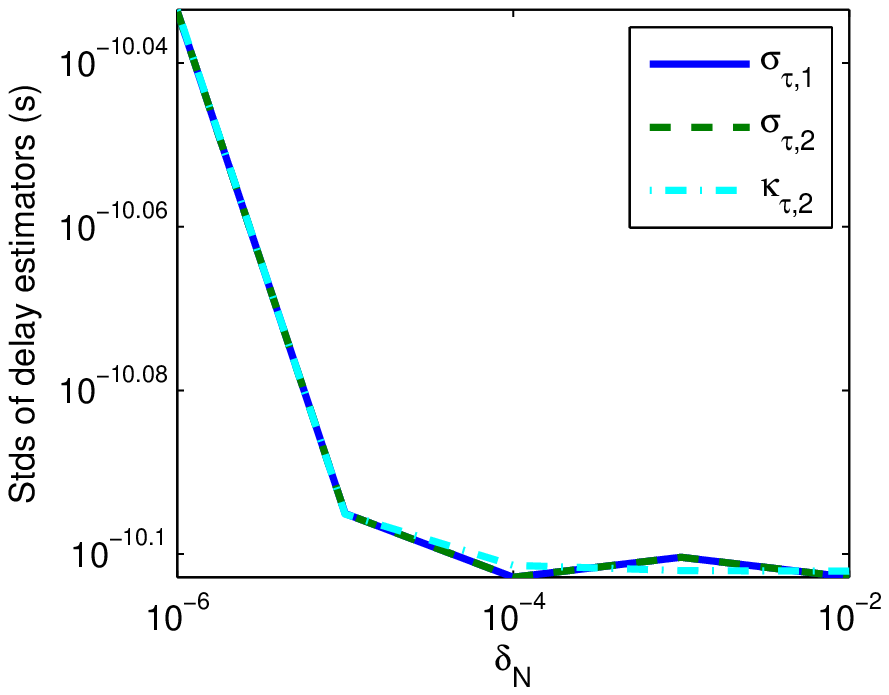} \label{tau_delN}} 
		\quad\subfigure[]{\includegraphics[width=6cm]{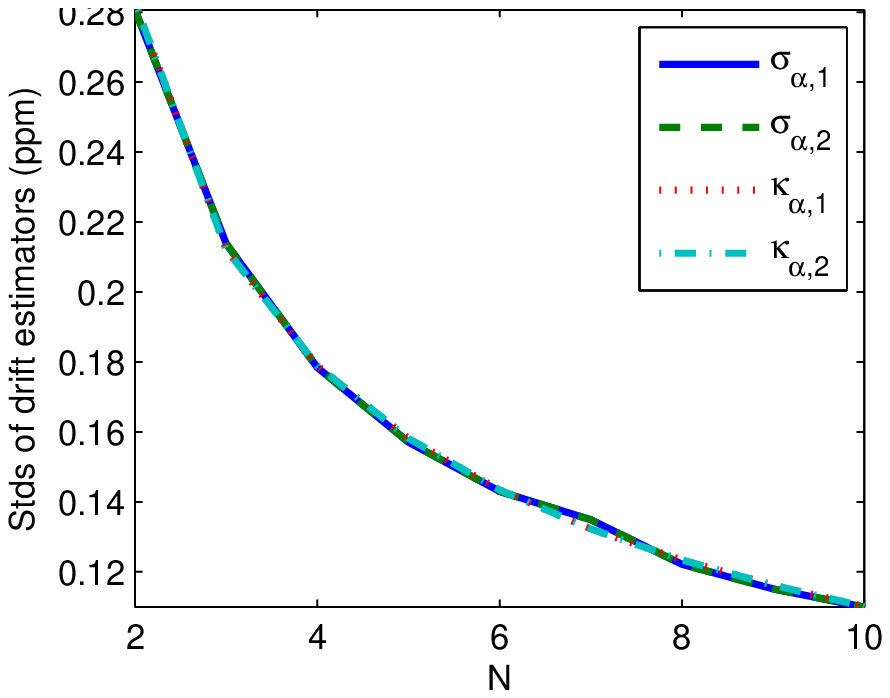} \label{alpha_N}} \hspace{-0.045\textwidth}
		\quad\subfigure[]{\includegraphics[width=6cm]{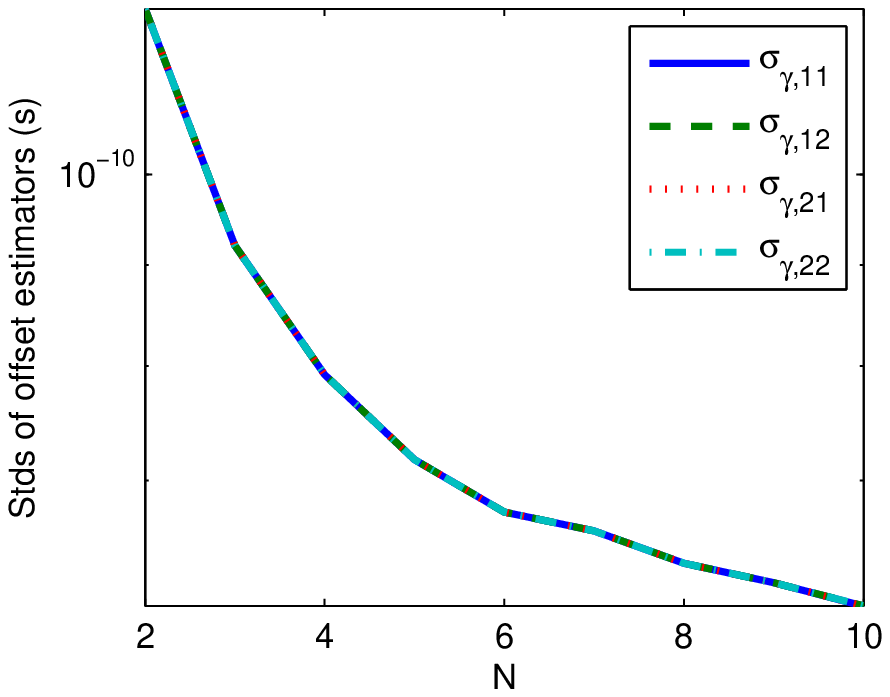} \label{gamma_N}} \hspace{-0.045\textwidth}
		\quad\subfigure[]{\includegraphics[width=6cm]{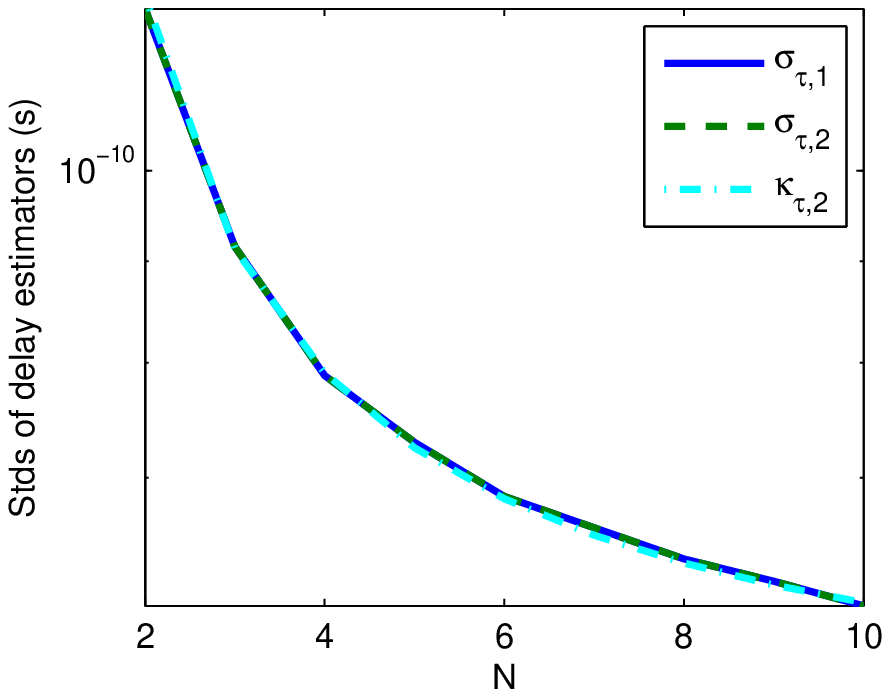} \label{tau_N}} 
	\caption{(a)--(l) Stds for drift, offset and delay estimation, respectively, w.r.t. $\sigma_A$, $\sigma_R$, $\delta_N$ and $N$, respectively.}
	\end{center}
\end{figure*}

Fig. \ref{alpha_sigA} shows that $\hat{\alpha}_1$ and $\hat{\alpha}_2$ achieve the same performance; they both achieve the CRLB which is independent of $\sigma_A$. The variance of an unbiased estimator can never be lower than the CRLB. However, $\sigma_{\alpha,1}$, $\sigma_{\alpha,2}$ are sometimes lower than $\kappa_{\alpha,2}$ because they are obtained by simulation.
Fig. \ref{gamma_sigA} shows that $\hat{\gamma}_1$ and $\hat{\gamma}_2$ approximately achieve the same performance. The achieved variances increase with $\sigma_A$. 
Fig. \ref{tau_sigA} shows that $\hat{\tau}_1$ and $\hat{\tau}_2$ approximately achieve the same performance. They both achieve the CRLB that increases with $\sigma_A$.

\subsection{Impact of $\sigma_R$}\label{sigR_subsec}

For the estimation of $\gamma$ and $\tau$, we observe in Figs. \ref{gamma_sigR} and \ref{tau_sigR} the same results discussed in Sec. \ref{sigA_subsec}. However, the variance achieved by the estimators of $\alpha$ increases now with $\sigma_R$ as can be observed in Fig. \ref{alpha_sigR}.



\subsection{Impact of $\delta_N$}\label{delN_subsec}

The variance achieved by the estimators of $\alpha$ decreases as $\delta_N$ increases as can be seen in Fig. \ref{alpha_delN}. This result can be expected from \eqref{alphaestn_eq}.

\smallskip

Fig. \ref{gamma_delN} shows that the variances achieved by the estimators of $\gamma$ decrease as $\delta_N$ increases, then increase to reach a given ceil. The convergence to a constant value is due to the fact that the lowest variance achieved by an estimator of $\gamma$ should be a function of $\sigma_A$, $\sigma_R$ and $N$. However, to understand the non-monotonous behavior of the achieved variance we need a closed-form expression of the variance or the CRLB.

\smallskip

We can see in Fig. \ref{tau_delN} that the variances achieved by the estimators of $\tau$ decrease as $\delta_N$ increases until they converge to a constant value. This result is expected like for the estimation of $\gamma$.



\subsection{Impact of $N$}\label{N_subsec}

We can observe in Figs. \ref{alpha_N}--\ref{tau_N} that the variances achieved by the estimators of $\alpha$, $\gamma$ and $\tau$ decrease as $N$ increases; in fact, by increasing $N$ we increase the total SNR because $\hat{t}_{R,1}',\cdots,\hat{t}_{R,N}'$ are independent.



\section{Conclusion}

We have considered the joint estimation of the time delay between two transceivers and the offset and the drift of an imperfect clock. 
We have proposed a protocol for the synchronization of the transceivers.
We have proposed some empirical estimators for the delay, the offset and the drift. Also, we have derived the CRLBs and the joint MLE of the delay and the drift.
We have studied the impact of the parameters of the protocol and the TOA estimation variance on the achieved performances. Some theoretical results are validated by simulation.


\footnotesize
\bibliographystyle{IEEEtran}
\bibliography{IEEEabrv,all_chapters_ref_abbrev}

\begin{thebibliography}{10}
\providecommand{\url}[1]{#1}
\csname url@samestyle\endcsname
\providecommand{\newblock}{\relax}
\providecommand{\bibinfo}[2]{#2}
\providecommand{\BIBentrySTDinterwordspacing}{\spaceskip=0pt\relax}
\providecommand{\BIBentryALTinterwordstretchfactor}{4}
\providecommand{\BIBentryALTinterwordspacing}{\spaceskip=\fontdimen2\font plus
\BIBentryALTinterwordstretchfactor\fontdimen3\font minus
  \fontdimen4\font\relax}
\providecommand{\BIBforeignlanguage}[2]{{%
\expandafter\ifx\csname l@#1\endcsname\relax
\typeout{** WARNING: IEEEtran.bst: No hyphenation pattern has been}%
\typeout{** loaded for the language `#1'. Using the pattern for}%
\typeout{** the default language instead.}%
\else
\language=\csname l@#1\endcsname
\fi
#2}}
\providecommand{\BIBdecl}{\relax}
\BIBdecl

\bibitem{fcc}
{Federal Communications Commission ({FCC})}, ``Revision of part 15 of the
  commission rules regarding ultra-wideband transmission systems,'' in
  \emph{FCC 02-48}, Apr. 2002.

\bibitem{ec1}
{Commission of the European Communities ({EC})}, ``Commission decision of
  21/ii/2007 on allowing the use of the radio spectrum for equipment using
  ultra-wideband technology in a harmonised manner in the community,'' Feb.
  2007.

\bibitem{ec2}
------, ``Final report from {CEPT} in response to {EC} mandates on the
  harmonized introduction of radio applications based on ultra-wideband ({UWB})
  technology,'' Mar. 2007.

\bibitem{Lee}
J.-Y. Lee and R.~Scholtz, ``Ranging in a dense multipath environment using an
  {UWB} radio link,'' \emph{IEEE Journal on Selected Areas in Communications},
  vol.~20, no.~9, pp. 1677--1683, Dec. 2002.

\bibitem{IEEE802.15.4a}
``{IEEE} {S}tandard for {I}nformation {T}echnology - {T}elecommunications and
  {I}nformation {E}xchange {B}etween {S}ystems - {L}ocal and {M}etropolitan
  {A}rea {N}etworks - {S}pecific {R}equirement {P}art 15.4: {W}ireless {M}edium
  {A}ccess {C}ontrol ({M}ac) and {P}hysical {L}ayer ({PHY}) {S}pecifications
  for {L}ow-{R}ate {W}ireless {P}ersonal {A}rea {N}etworks ({WPANs}),''
  \emph{IEEE Std 802.15.4a-2007 (Amendment to IEEE Std 802.15.4-2006)}, pp.
  1--203, 2007.

\bibitem{GeziciIEEE}
Z.~Sahinoglu and S.~Gezici, ``Ranging in the {IEEE} 802.15.4a standard,'' in
  \emph{IEEE Annual Wireless and Microwave Technology Conference (WAMICON
  2006)}, Dec. 2006, pp. 1--5.

\bibitem{dardari3}
D.~Dardari, A.~Conti, U.~Ferner, A.~Giorgetti, and M.~Win, ``Ranging with
  ultrawide bandwidth signals in multipath environments,'' \emph{Proc. IEEE},
  vol.~97, no.~2, pp. 404--426, Feb. 2009.

\bibitem{Kim}
H.~Kim, ``Double-sided two-way ranging algorithm to reduce ranging time,''
  \emph{IEEE Communications Letters}, vol.~13, no.~7, pp. 486--488, July 2009.

\bibitem{Xing}
L.~J. Xing, L.~Zhiwei, and F.~Shin, ``Symmetric double side two way ranging
  with unequal reply time,'' in \emph{IEEE Vehicular Technology Conference (VTC
  2007)}, Sept. 2007, pp. 1980--1983.

\bibitem{Jiang}
Y.~Jiang and V.~Leung, ``An asymmetric double sided two-way ranging for crystal
  offset,'' in \emph{International Symposium on Signals, Systems and
  Electronics (ISSSE 2007)}, July 2007, pp. 525--528.

\bibitem{Denis}
B.~Denis, J.-B. Pierrot, and C.~Abou-Rjeily, ``Joint distributed
  synchronization and positioning in {UWB} ad hoc networks using {TOA},''
  \emph{IEEE Transactions on Microwave Theory and Techniques}, vol.~54, no.~4,
  pp. 1896--1911, June 2006.

\bibitem{Shimizu}
Y.~Shimizu and Y.~Sanada, ``Accuracy of relative distance measurement with
  ultra wideband system,'' in \emph{2003 IEEE Conference on Ultra Wideband
  Systems and Technologies}, Nov. 2003, pp. 374--378.

\bibitem{Sivrikaya}
F.~Sivrikaya and B.~Yener, ``Time synchronization in sensor networks: a
  survey,'' \emph{IEEE Network}, vol.~18, no.~4, pp. 45--50, July 2004.

\bibitem{Zhen}
B.~Zhen, H.-B. Li, and R.~Kohno, ``Clock management in ultra-wideband
  ranging,'' in \emph{2007 Mobile and Wireless Communications Summit}, July
  2007, pp. 1--5.

\bibitem{mcaulay3}
R.~McAulay and D.~Sakrison, ``A {PPM}/{PM} hybrid modulation system,''
  \emph{IEEE Trans. Commun. Technol.}, vol.~17, no.~4, pp. 458--469, Aug. 1969.

\bibitem{mallat1}
A.~Mallat, J.~Louveaux, and L.~Vandendorpe, ``{UWB} based positioning in
  multipath channels: {CRB}s for {AOA} and for hybrid {TOA}-{AOA} based
  methods,'' in \emph{IEEE Int. Conf. Commun. (ICC 2007)}, June 2007, pp.
  5775--5780.

\bibitem{mallat2009mpc}
A.~Mallat, C.~Oestges, and L.~Vandendorpe, ``{CRB}s for {UWB} multipath channel
  estimation: Impact of the overlapping between the {MPC}s on {MPC} gain and
  {TOA} estimation,'' in \emph{IEEE Int. Conf. Commun. (ICC 2009)}, June 2009,
  pp. 1--6.

\bibitem{kay}
S.~Kay, \emph{Fundamentals of Statistical Signal Processing Estimation
  Theory}.\hskip 1em plus 0.5em minus 0.4em\relax Prentice-Hall, 1993.

\bibitem{Marsaglia1964}
G.~Marsaglia, ``Ratios of normal variables and ratios of sums of uniform
  variables,'' \emph{J. Amer. Statist. Assoc.}, vol.~60, no. 309, pp. 193--204,
  Mar. 1965.

\bibitem{Marsaglia2006}
------, ``Ratios of normal variables,'' \emph{J. Stat. Softw.}, vol.~16, no.~4,
  May 2006.

\end{thebibliography}


\end{document}